\documentclass[useAMS,usenatbib,a4paper,preprint]{mn2e}
\usepackage{graphicx}
\usepackage{amsmath}
\usepackage{txfonts}
\usepackage{mathrsfs}

\usepackage{color}

\def\aap{A\&A}
\def\apjl{ApJ}

\def\apjs{ApJS}

\def\apjs{ApJ Supplements}

\def\apjl{ApJ Letters}

\def\aap{A\&A}

\newcommand{\be}{\begin{equation}}
\newcommand{\ee}{\end{equation}}
\newcommand{\bea}{\begin{eqnarray}}
\newcommand{\eea}{\end{eqnarray}}

\title[The {\it JWST} High-redshift Galaxies]{The Cosmic Timeline Implied by the {\it JWST} High-redshift Galaxies}
\author[Fulvio Melia]{Fulvio Melia\thanks{John Woodruff Simpson
Fellow. E-mail: fmelia@email.arizona.edu}\\
Department of Physics, The Applied Math Program, and Department of Astronomy,
The University of Arizona, AZ 85721, USA}

\begin{document}

\date{}

\pagerange{\pageref{firstpage}--\pageref{lastpage}} \pubyear{2023}

\maketitle

\label{firstpage}

\begin{abstract}
The so-called `impossibly early galaxy' problem, first identified via the {\it Hubble Space
Telescope}'s observation of galaxies at redshifts $z> 10$, appears to have been exacerbated 
by the more recent {\it James Webb Space Telescope} ({\it JWST}) discovery of galaxy candidates 
at even higher redshifts ($z\sim 17$) which, however, are yet to be confirmed spectroscopically. 
These candidates would have emerged only $\sim 230$ million years after the big 
bang in the context of $\Lambda$CDM, requiring a more rapid star formation in the earliest 
galaxies than appears to be permitted by simulations adopting the concordance model 
parameters. This time-compression problem would therefore be inconsistent with the age-redshift 
relation predicted by $\Lambda$CDM. Instead, the sequence of star formation and galaxy assembly 
would confirm the timeline predicted by the $R_{\rm h}=ct$ universe, a theoretically advanced 
version of $\Lambda$CDM that incorporates the `zero active mass' condition from general 
relativity. This model has accounted for many cosmological data better than 
$\Lambda$CDM, and eliminates all of its inconsistencies, including the horizon and initial
entropy problems. The latest {\it JWST} discoveries at $z\gtrsim 14$, if confirmed, 
would add further support to the idea that the $R_{\rm h}=ct$ universe is favored by the 
observation over the current standard model.
\end{abstract}


\section{Introduction}\label{intro}
A surprising number of high-redshift galaxy candidates ($z> 12$) have already been
discovered by the {\it James Webb Space Telescope} ({\it JWST}) in just the first few weeks of
operation (see Table~\ref{tab1}). Identified through the Early Release Observations
(ERO) \citep{Pontoppidan:2022}, the Cosmic Evolution Early Release Science (CEERS)
\citep{Finkelstein:2022} and Through the Looking GLASS (GLASS-{\it JWST}) \citep{Treu:2022} science
programs, many of them surpass the distance record previously set at $z=11.1$ by 
the {\it Hubble Space Telescope} ({\it HST}) \citep{Oesch:2016}. This is certainly true
of candidates up to $z\sim 13$, whose redshift has been confirmed spectroscopically
\citep{Robertson:2022}.

But the fact that some of these well-formed $\sim 10^9\,M_\odot$ structures (at $z\sim 
16-17$) appear to have emerged only $\sim 230$ Myr after the big bang contrasts with 
their predicted formation in the standard model of cosmology, which we here take to be 
$\Lambda$CDM with the {\it Planck} optimized parameters: a Hubble constant, $H_0=67.4
\pm0.5$ km s$^{-1}$ Mpc$^{-1}$, a matter density $\Omega_{\rm m}=0.315\pm0.007$, scaled 
to today's critical density ($\equiv 3c^2H_0^2/ 8\pi G$), and a spatial curvature constant, 
$k\approx 0$ \citep{PlanckVI:2020}. In discussing this `impossibly early galaxy' 
problem \citep{Melia:2014a,Melia:2020}, two principal issues typically emerge. The first 
is whether the gas budget in the early Universe, notably the fraction of baryons 
condensed within an assumed dark-matter halo distribution, was sufficient to account 
for this high-$z$ galaxy demographic \citep{Behroozi:2018}. The answer could be yes 
\citep{Donnan:2022}, as long as all of the available baryonic gas in halos was converted 
into stars. The $z\sim 16-17$ galaxy candidates fall close to the $\Lambda$CDM limit, but 
do not exceed it.

\begin{table*}
\begin{center}
\begin{minipage}{580pt}
\caption{{\it JWST} highest-redshift galaxies and their derived properties}\label{tab1}%
{\footnotesize
\begin{tabular}{@{}lllrcl@{}}
\hline\hline \\
Name & $\quad\;\; z^a$  & $\log(M/M_\odot)$ & SFR$\quad\;$ & Stellar Age$^b$ & Reference \\
     &        &                   & ($M_\odot$ yr$^{-1}$) & (Myr) &   \\ \\
\hline \\
1. S5-z17-1        & $16.66^{+1.86}_{-0.34}$ & $\;8.8^{+0.8}_{-0.5}$     & $9.7^{+30.7}_{-6.2}$ &
-                & \citet{Harikane:2022}   \\
2. CEERS-93316     & $16.4^{+0.1}_{-0.1}$    & $\;9.0^{+0.4}_{-0.5}$     & $10.0^{+10}_{-6.8}\,$ &
$20^{+40}_{-10}$ & \citet{Donnan:2022,Naidu:2022b,Harikane:2022}  \\
3. S5-z12-1        & $13.72^{+0.86}_{-1.92}$ & $\;8.1^{+1.3}_{-0.3}$     & $2.2^{+15.5}_{-1.0}\,$ &
-               & \citet{Harikane:2022} \\
4. WHL0137-5021    & $12.8^{+1.1}_{-12.5}$   & $\;8.53^{+0.18}_{-0.32}$  & $5.1^{+1.9}_{-1.1}\;\;$ &
$58^{+35}_{-35}$ & \citet{Bradley:2022}  \\
5. WHL0137-5124    & $12.8^{+1.9}_{-12.4}$   & $\;8.65^{+0.20}_{-0.30}$  & $6.9^{+3.2}_{-1.9}\;\;$ &
$59^{+35}_{-37}$ & \citet{Bradley:2022}  \\
6. GLASS-z13          & $12.4\pm{0.2}$   & $\;9.0^{+0.3}_{-0.4}$  & $7^{+4}_{-3}\;\,$ &
$71^{+32}_{-33}$ & \citet{Naidu:2022a,Harikane:2022}  \\
7. GLASS-z12-1        & $12.22^{+0.04}_{-0.11}$ & $\;8.6^{+0.8}_{-0.4}$ & $3.0^{+11.3}_{-0.6}\;$ &
- & \citet{Harikane:2022,Donnan:2022}  \\
8. Maisie's Galaxy    & $11.8^{+0.2}_{-0.3}$   & $\;8.5^{+0.29}_{-0.44}$ & $2.1^{+4.8}_{-2.0}\;\;$ &
$18^{+18}_{-9}$  & \citet{Finkelstein:2022,Harikane:2022}  \\
9. GN-z11$^c$ & $11.09^{+0.08}_{-0.12}$ & $\;9.0\pm{0.4}$ & $24\pm{10}\;\,$ &
$40^{+60}_{-34}$   & \citet{Oesch:2016} \\
10. GLASS-z11          & $10.6\pm{0.3}$   & $\;9.4\pm{0.3}$  & $12^{+9}_{-4}\;\;$ &
$111^{+43}_{-54}\;\,$ & \citet{Naidu:2022a,Harikane:2022,Donnan:2022}  \\
11. WHL0137-3407    & $10.5^{+1.0}_{-10.5}$   & $\;8.78^{+0.17}_{-0.33}$  & $7.3^{+2.5}_{-1.2}\;\,$ &
$70^{+50}_{-44}$ & \citet{Bradley:2022}  \\
12. WHL0137-5347    & $10.2^{+0.9}_{-9.7}$   & $\;9.01^{+0.21}_{-0.37}$  & $14.6^{+5.8}_{-3.5}\;\,$ &
$62^{+54}_{-43}$ & \citet{Bradley:2022}  \\
13. WHL0137-5330    & $10.0^{+1.1}_{-7.9}$   & $\;8.77^{+0.15}_{-0.26}$  & $6.4^{+2.6}_{-1.8}\;\,$ &
$83^{+52}_{-48}$ & \citet{Bradley:2022}  \\ \\
\hline
\end{tabular}
}
\footnotetext[1]{Photometric redshift (with $2\sigma$ uncertainties) for the WHL sources calculated
using the \citet{Calzetti:2000} dust law. GLASS, \newline\hangindent15pt CEERS and Maisie's Galaxy
redshifts (with $1\sigma$ uncertainties) were fit with the EAZY code \citep{Brammer:2008}.}
\footnotetext[2]{Mass-weighted age of the current luminous stars, when available.}
\footnotetext[3]{Discovered by {\it HST} prior to {\it JWST}.}
\end{minipage}
\end{center}
\end{table*}

The second concerns whether the dynamics of structure formation could account for the 
highly compressed timeline implied by these discoveries (for the most recent work on this topic, see
\citealt{Yajima:2022,Keller:2022,Kannan:2022,Inayoshi:2022,Haslbauer:2022,Mirocha:2023,Whitler:2023}).
It is the dynamics, of course, coupled to the physical processes responsible for cooling the gas, 
that would have governed how quickly stars could condense and assemble into billion solar-mass
structures. One must also fold into this discussion how the `stellar age' of the high-$z$
sources (column~5 in Table~\ref{tab1}) should be interpreted. An examination of the galaxy
age versus star formation activity at $z> 8$ \citep{Furtak:2023,Whitler:2023} suggests that 
the young stellar populations producing much of the current luminosity are built upon older 
components that formed at $z> 15$, and are being observed during bursts of star formation.

\section{High-$z$ Galaxies in $\Lambda$CDM}\label{galaxies}
A more indicative evolutionary history for these galaxies is therefore provided by the broad 
range of simulations tracing the growth of initial perturbations consistent with the measured 
anisotropies in the cosmic microwave background. This study has evolved considerably 
over the past decade, as each new set of observations has pushed the formation of galaxies to 
progressively higher redshifts. The first generation of simulations
\citep{Barkana:2001,Miralda:2003,Bromm:2004,Ciardi:2005,Glover:2005,Greif:2007,Wise:2008,Salvaterra:2011,Greif:2012,Jaacks:2012} 
began to elucidate how the first (Pop~III) stars probably formed by redshift $z\sim 20$, nestled 
in the core of dark-matter halos with mass $M_{\rm halo}\sim 10^6\,M_\odot$ 
\citep{Haiman:1996,Tegmark:1997,Abel:2002,Bromm:2002}. This delay after the big bang resulted 
from the combined influence of several processes, including the initial gravitational collapse
of the dark-matter perturbations and the subsequent inefficient, radiative cooling of the
primordial gas. The baryonic matter cooled and condensed into stars only after enough molecular 
hydrogen accumulated to enhance the energy loss rate \citep{Galli:1998,Omukai:1998}. In the 
standard model, the Universe would have been $\sim 180$ Myr old at redshift 20. But not all
of the halos and their baryonic content would necessarily have taken this long to condense.
The more recent simulations \citep{Yajima:2022,Keller:2022,Kannan:2022,Inayoshi:2022,Haslbauer:2022,Mirocha:2023,Whitler:2023},
in particular, show that the halos could have been distributed across this age, some appearing 
perhaps as early as $\sim 120$ Myr after the big bang (more on this below).

A typical baryonic gas cloud could subsequently have formed a protostar at the 
center of its halo host, eventually growing to become a $> 100\,M_\odot$ main sequence 
(Pop~III) star \citep{Kroupa:2002,Chabrier:2003}. This is where a second major difficulty 
emerges, however. The UV radiation from such massive stars would have destroyed all 
of the $H_2$ in the original condensation, suggesting that most of the minihalos 
likely contained---at most---only a handful of Pop~III stars \citep{Yoshida:2008}. Thus, many of
these early structures were probably not the galaxies we see today. Nevertheless, a sufficiently
broad distribution of Pop~III masses could have included many below $\sim 100\,M_\odot$, whose 
impact on the star formation would have been less severe
\citep{Yajima:2022,Keller:2022,Kannan:2022,Inayoshi:2022,Haslbauer:2022,Mirocha:2023,Whitler:2023}.

The outpouring of radiative and mechanical energy from this first phase of stellar formation
would have reheated and expelled the surrounding gas, further delaying the formation of additional
stars until the plasma had time to cool again and condense to high densities. The time for this 
gas re-incorporation would have been another $\sim 100$ Myr, i.e., roughly the dynamical time 
for a first-galaxy halo to assemble \citep{Yoshida:2004,Johnson:2007}. And now the Universe was 
almost 300 Myr old.

To its credit, this scenario does provide a plausible explanation for the size of the
observed galaxies. It suggests that a criticial mass of $> 10^8\,M_\odot$, along with an
implied virial temperature $> 10^4$ K, would have allowed atomic line emission to cool
the condensing gas \citep{Wise:2007}, and recollect and mix all components of the previously
shocked plasma \citep{Greif:2007}. In other words, at least the mass inferred for the earliest
{\it JWST} galaxies (Table~\ref{tab1}) is consistent with this theoretical understanding.

But in the context of these simulations, there's no getting around the fact that 
the appearance of $\sim 10^9\,M_\odot$ galaxies at $z\sim 16-17$, if confirmed spectroscopically, 
would create a significant problem. One needs a billion new stars to have formed in only 
$\sim 70-90$ Myr by the time the Universe was only $t\sim 230$ Myr old. None of the calculations 
to date have been able to account for the formation of galaxies at this redshift, when the 
expelled, hot gas hadn't re-cooled and re-condensed yet.

This tension between the theoretical and observational timelines has motivated the introducion 
of additional features and physical processes designed to mitigate the disagreement as much as 
possible \citep{Yajima:2022,Keller:2022,Kannan:2022,Inayoshi:2022,Haslbauer:2022,Mirocha:2023,Whitler:2023}.
For example, in their most detailed simulations to date, \cite{Yajima:2022} and \cite{Keller:2022} 
have demonstrated that the large scatter in cooling times and the presence of systems with weaker 
Pop~III supernovae that expel far less of the condensed baryonic gas \citep{Kitayama:2005,Frebel:2015} 
would have allowed galaxies observed by {\it JWST} at $z\lesssim 14$ to still have formed in the 
context of $\Lambda$CDM.

\begin{figure}
\centering
\includegraphics[width=0.48\textwidth]{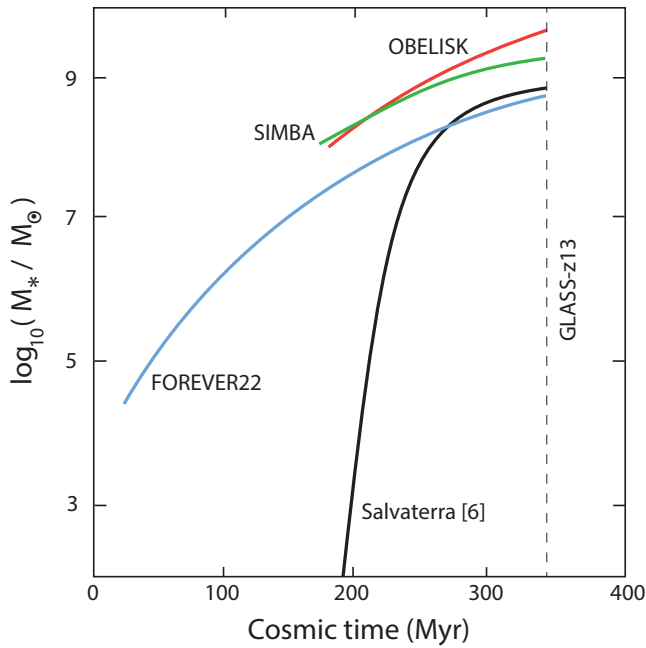}
\caption{Comparison of galaxy growth curves with four different simulations for GLASS-z13 
(line 6 in Table~1), with a final stellar mass $M_*\sim 10^9\;M_\odot$: Salvaterra [6]
corresponding to trajectory [6] in Figure~2; FOREVER22 \citep{Yajima:2022}; SIMBA \citep{Keller:2022}; 
and OBELISK \citep{Keller:2022}. The vertical dashed line indicates the age of this galaxy
($\sim 345$ Myr) in the context of {\it Planck}-$\Lambda$CDM.}
\end{figure}\label{fig1}

\cite{Kannan:2022} have shown that a variable stellar initial mass function may also have produced 
some galaxies earlier than previously thought. A top-heavy stellar mass distribution appears to have
a similar effect \citep{Inayoshi:2022}, while different star formation histories could reduce the 
actual stellar masses of the galaxy candidates, thereby partially alleviating the tension 
\citep{Haslbauer:2022}. \cite{Mirocha:2023} suggest that at least three modifications to the 
{\it HST}-calibrated models would help lessen the tension: (i) the adoption of halo 
mass-independent star formation (SFR) efficiencies; (ii) a substantial scatter in galaxy SFRs 
at fixed halo masses; and (iii) the non-trivial effects of dust, both on the inferred {\it JWST} 
colours and on the produced stellar masses and ages. Finally, \cite{Whitler:2023} conclude that 
the tension may be eased if young stellar populations formed in these early galaxies on top of 
older stellar populations.

Nevertheless, even with all of these modifications, the predicted galaxy masses at $z\lesssim 14$ 
appear to fall short of those observed by factors of a few. And they are significantly smaller 
than those of the galaxy candidates at $z\sim 16-17$. Thus, if the {\it JWST} highest redshift 
sources are eventually confirmed, even the more optimistic recent simulations would be unable 
to explain their origin.

The four galaxy growth curves in Figure~1 compare the time required to reach 
the inferred stellar mass of GLASS-z13 (the sixth entry in Table~1) by $t\sim 345$ Myr
(corresponding to its redshift $z\sim 12.4$), based on four of the main simulations
discussed above. For the sake of giving $\Lambda$CDM the most optimal evolutionary
outcome, we shall adopt the \cite{Salvaterra:2011} and \cite{Jaacks:2012} frameworks 
which, though not as detailed and nuanced as the more recent work, predict a shorter 
growth time once star formation is initiated, thus making it easier to fit the observed 
galaxies within the $\Lambda$CDM timeline. 

\begin{figure}
\centering
\includegraphics[width=0.48\textwidth]{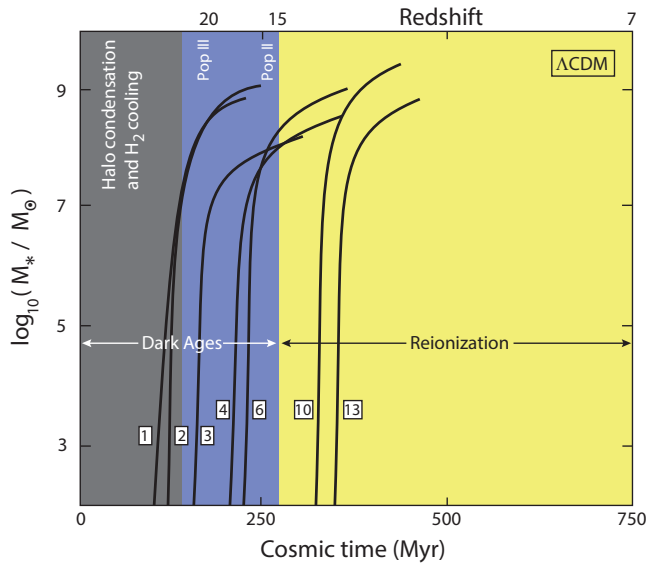}
\caption{Growth of stellar mass in the high-$z$ galaxy candidates discovered 
by {\it JWST} at $10< z< 16$,
as a function of cosmic time $t$, in $\Lambda$CDM. The principal epochs are (i) the initial
halo condensation and cooling due to molecular hydrogen. This epoch typically extended over
the period $0.4\lesssim t\lesssim 180$ Myr, but could have been as short as $\sim 120$ Myr
for some of the objects; (ii) the formation of the first Pop~III stars at $t\sim 120-180$ 
Myr, i.e., $z\gtrsim 20$ in this model, (iii) the transition to Pop~II star formation at 
$t\lesssim 280$ Myr, and the observed Epoch of Reionization (EoR) from $z\sim 15$ down to 
$6$ (i.e., $280< t< 927$ Myr). These galaxy growth trajectories are primarily based on the 
observed SFRs and the hydrodynamical simulations in \citet{Jaacks:2012}, cross-checked with 
independent and alternative calculations in \citet{Salvaterra:2011}. The labels on the curves 
correspond to the catalog listings in Table~\ref{tab1}.}
\end{figure}\label{fig2}

The various factors discussed above may now be seen more quantitatively with the 
simulated galaxy growth trajectories shown in Figure~2, which also includes several 
critical epochs in $\Lambda$CDM. We assume that the {\it JWST} galaxy candidates in Table~\ref{tab1}
followed a history of growth like those of \cite{Salvaterra:2011} and \cite{Jaacks:2012} at
$z\sim 8$, except that their cosmic time $t$ is suitably translated to match the redshift at 
which they are observed. This is justified by the fact that $t$ is actually the proper time in 
the comoving frame, and the Birkhoff theorem ensures that the local growth rate was not overly
affected by the cosmic expansion exterior to the bound system \citep{Weinberg:1972,Melia:2020}.
In other words, once a galaxy halo becomes gravitationally bound and star formation is initiated, 
its evolution thereafter should be roughly translationally invariant in $t$.

In the \cite{Salvaterra:2011} simulations, the doubling time (i.e., the inverse of the specific 
star formation rate sSFR, defined as the stellar mass created per unit time per billion 
solar-masses) and the evolutionary time at which the galaxy is observed, appears to be
universally equal to $\sim 0.1-0.3$. By comparison, the \cite{Jaacks:2012} calculations
show that the SFR for high-$z$ galaxies is `bursty,' with an average value between 
$z\sim 15$ and $z\sim 6$ following an exponentially increasing function with characteristic 
timescale $t_{\rm c}\sim 70-200$ Myr, scaling with stellar mass in the range 
$10^6< M_*< 10^{10}\,M_\odot$. The trajectories plotted in Figure~2 follow this exponential 
growth, using the observed galaxy mass and redshift to fix the end points. Given that the sSFRs 
probably fluctuated during their evolution \citep{Furtak:2023,Whitler:2023}, we take an 
average of the star formation rates quoted in Table~\ref{tab1}, i.e., 
$\langle{\rm sSFR}\rangle\sim 12.4$ Gyr$^{-1}$, as a fiducial value for each galaxy.

The overall impression one gets from this illustration is that the new 
{\it JWST} high-$z$ galaxies, if confirmed, would not be consistent with the standard picture 
in $\Lambda$CDM. The previously growing tension developing at $z\sim 12$ has now become
a more serious discordance at $z\sim 16-17$. There does not appear to be any possibility
with our current physical theories of explaining how a billion-solar mass aggregate of stars
could have condensed even before the primoridal gas was allowed to cool and form most of the 
very first Pop~III, and any of the Pop~II, populations. In this regard,
our conclusion concerning the implausibility of forming the {\it JWST} high-$z$
galaxies in the context of $\Lambda$CDM is fully consistent with the findings of
an alternative approach to this problem \citep{Boylan:2022}, based on the use of
the Sheth-Tormen \citep{Sheth:1999} mass function to determine the abundance of
massive halos at high redshifts \citep{Wang:2022}.

\section{The Timeline in $R_{\rm h}=ct$}\label{Rhct}
But previous comparative tests between $\Lambda$CDM and a theoretically more advanced
version, known as the $R_{\rm h}=ct$ universe \citep{MeliaShevchuk:2012,Melia:2020},
have already hinted at the possibility that the age-redshift relation in the latter
may be a better match to the Universe's evolutionary history than that 
in the former. For example, $\Lambda$CDM has considerable difficulty accounting for 
the seeding and growth of billion-solar mass black holes by redshift $z\sim 7-8$, 
while the timeline in $R_{\rm h}=ct$ matches them very well \citep{Melia:2013b,Melia:2018c}. 
A more complete description of this model, including its theoretical
foundation and a comparison of its predictions with the data, may be seen in 
\citet{Melia:2018e,Melia:2020}. A review of the current problems with the standard 
model, pointing to a need for further development, e.g., as suggested by $R_{\rm h}=ct$, 
is provided in \citet{Melia:2022e}.

\begin{figure}
\centering
\includegraphics[width=0.48\textwidth]{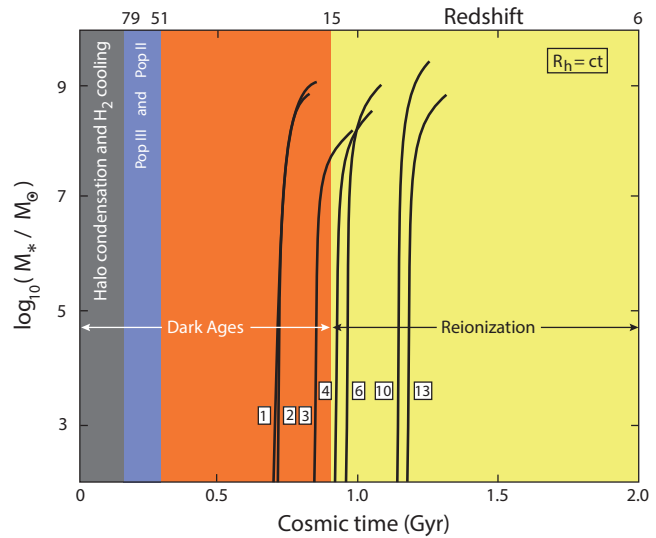}
\caption{Same as Figure~2, except now for the $R_{\rm h}=ct$ Universe. The EoR
here corresponds to $906\;{\rm Myr}< t< 2.07$ Gyr, and the Dark Ages extend
up to $\sim 906$ Myr. The first Pop~III stars emerged at $z\gtrsim 79$ 
and the transition to Pop~II stars occurred at $z\sim 51$. The extended period 
between the onset of Pop~II star formation and the appearance of the first 
{\it JWST} galaxies (shown here in red) is absent in Figure~2. In this cosmology, 
the {\it JWST} galaxy candidates are seen towards the end of the dark ages, where 
one would expect them to be if they were responsible for re-ionizing the intergalactic 
medium. Most importantly, all of these primordial galaxies would have started their 
growth {\it well after} the transition from Pop~III to Pop~II star 
formation had been completed at $\sim 280$ Myr.}
\end{figure}\label{fig3}

One of the essential features of $R_{\rm h}=ct$ that distinguishes it from $\Lambda$CDM
is its expansion factor, $a(t)\propto t$, which results in the simple age-redshift relation
$1+z={t_0/t}$ in terms of the current age, $t_0$ of the Universe. In this cosmology, the
gravitational radius $R_{\rm h}$ is equivalent to the Hubble radius $c/H(t)$ \citep{Melia:2018b},
so $t_0={1/ H_0}$. Thus, if for simplicity we use the same Hubble constant as $\Lambda$CDM,
we find that $t_0\approx 14.5$ Gyr. The {\it JWST} galaxy trajectories recalculated
with these relations are displayed in Figure~3, along with the correspondingly
adjusted temporal phases. The EoR redshift range $6< z< 15$ here corresponds
to $906\;{\rm Myr}< t < 2.07$ Gyr. That is, the Dark Ages ended at
$t\sim 906$ Myr, providing ample time for the Universe to assemble billion-solar mass
galaxies once Pop~III and Pop~II stars started forming in numbers. Very tellingly, all
of the high-$z$ galaxy candidates discovered so far appear towards the end of the dark
ages, where one would expect them to be if they contributed---perhaps even dominated---the
reionization process. In this model, one would need to find a billion-solar mass galaxy
at $z\sim 50$ to run into a similar age-redshift inconsistency as that in $\Lambda$CDM.

\section{Conclusion}\label{conclusion}
The time compression problem in the standard model has been worsening for several years.
Attempts at remedying the situation with the premature formation of supermassive black holes
have focused on two principal modifications: (i) the creation of massive (i.e.,
$\sim 10^5\,M_\odot$) seeds; and (ii) super-Eddington accretion rates. The first of these 
is still speculative because it requires the collapse of an essentially zero angular momentum, 
optically-thick, radiation-dominated plasma, which would have experienced substantial support 
from its internal pressure \citep{Melia:2009}; the second appears to have been ruled out by 
measurements suggesting that the most distant quasars are accreting at or below their Eddington 
rate \citep{Mortlock:2011,DeRosa:2011,Willott:2010a}.

The {\it JWST} discovery of high-$z$ galaxy candidates may have worsened this
timing problem considerably if their redshifts are confirmed spectroscopically, because 
billion-solar mass structures must have formed in only $\sim 70-90$ Myr in some cases, and 
even prior to formation of the very first stars in others. The simulations completed to 
date in the context of $\Lambda$CDM have difficulty accounting for this outcome at 
$z\sim 17$, reinforcing the view that the standard model may not be able 
to account for the formation of structure at cosmic dawn, if the redshifts of these 
candidate galaxies are confirmed spectroscopically.

Instead, the timeline predicted by the $R_{\rm h}=ct$ cosmology would fit the birth
and growth of both high-$z$ quasars and galaxies very well, adding to the growing
body of evidence supporting the introduction of the zero active mass condition
from general relativity as an indispensible modification to $\Lambda$CDM.

Of course, there are still at least two ways out of this dilemma. First, the 
{\it JWST} candidate galaxies at $z\gtrsim 14$ may simply be mis-identified sources at lower 
redshifts. According to \cite{Furtak:2023}, only about half of the high-$z$ galaxy
photometric redshifts may be safely ruled out as low-$z$ interlopers. The other half,
including those of the most distant candidates, still await spectroscopic confirmation. 
Second, it is possible that we may be missing something in the basic theory, and
this caveat cannot be ignored. The initial cooling of the primordial gas may have
been due to something other than molecular hydrogen. An unknown process 
may have permitted the plasma to cool more efficiently, allowing Pop~III stars to form
even earlier than $t\sim 120-180$ Myr. 

Certainly, the most recent simulations of \cite{Yajima:2022} and \cite{Keller:2022} 
indicate that such a process permitting the formation of Pop III stars as early as 
$\lesssim 100$ Myr would lessen the tension with the $z\sim 16-17$ galaxies in $\Lambda$CDM. 
This may not completely resolve the problem, but would go a long way to mitigating the 
overall disagreement between the {\it JWST} observations and the standard model. 
Future simulations will probe such possibilities in even greater detail, perhaps 
uncovering a solution based on new physics in $\Lambda$CDM. As of today, however, 
the {\it JWST} discoveries---if confirmed---would support the timeline in 
$R_{\rm h}=ct$, but not in $\Lambda$CDM.

\section*{Acknowledgments}
I am very grateful to the anonymous referee for a detailed and very 
constructive review, which has led to significant improvements in the manuscript. 

\section*{DATA AVAILABILITY STATEMENT}
No new data were generated or analysed in support of this research.

\bibliographystyle{mn2e.bst}
\bibliography{ms.bib}

\label{lastpage}
\end{document}